\documentclass[%
 reprint,
 amsmath,amssymb,
 aps,
prd,
]{revtex4-1}

\usepackage{graphicx}
\usepackage{dcolumn}
\usepackage{bm}
\usepackage{slashed}
\usepackage{color}
\usepackage{textcomp}

\newcommand{\aref}[1]{App.~\ref{#1}}
\newcommand{\eq}[1]{Eq.~\eqref{#1}}
\newcommand{\seqs}[1]{Eqs.~\eqref{#1}}
\newcommand{\eqs}[2]{Eqs.~\eqref{#1} and \eqref{#2}}
\newcommand{\fref}[1]{Fig.~\ref{#1}}

\newcommand{\dd}{\mathrm{d}}
\newcommand{\fand}{\text{ and }}
\newcommand{\bs}{\boldsymbol}
\newcommand{\nn}{\nonumber}
\newcommand{\wt}{\widetilde}

\def\mnras{Mon. Not. R. Astron. Soc. }

\def\apj{Astrophys. J.}
\def\apjl{Astrophys. J. Lett.}
\def\prl{Phys. Rev. Lett.}

\begin{document}

\preprint{APS/123-QED}

\title{Detecting the Beaming Effect of Gravitational Waves}

\author{Alejandro Torres-Orjuela}
\affiliation{Astronomy Department, School of Physics, Peking University, 100871 Beijing, China}
\affiliation{Kavli Institute for Astronomy and Astrophysics at Peking University, 100871 Beijing, China}

\author{Xian Chen}
\email{Corresponding author: xian.chen@pku.edu.cn}
\affiliation{Astronomy Department, School of Physics, Peking University, 100871 Beijing, China}
\affiliation{Kavli Institute for Astronomy and Astrophysics at Peking University, 100871 Beijing, China}

\author{Zhoujian Cao}
\affiliation{Institute of Applied Mathematics, Academy of Mathematics and Systems Science, Chinese Academy of Sciences, Beijing 100190, China}

\author{Pau Amaro-Seoane}
\affiliation{Institute of Space Sciences (ICE, CSIC) \& Institut d'Estudis Espacials de Catalunya (IEEC) at Campus UAB, 08193 Barcelona, Spain}
\affiliation{Institute of Applied Mathematics, Academy of Mathematics and Systems Science, Chinese Academy of Sciences, Beijing 100190, China}
\affiliation{Kavli Institute for Astronomy and Astrophysics at Peking University, 100871 Beijing, China}
\affiliation{Zentrum f{\"u}r Astronomie und Astrophysik, TU Berlin, 10623 Berlin, Germany
}

\author{Peng Peng}
\affiliation{Astronomy Department, School of Physics, Peking University, 100871 Beijing, China}

\date{\today}

\begin{abstract}

The models currently used in the detection of gravitational waves (GWs) either
do not consider a relative motion between the center-of-mass of the source and
the observer, or usually only consider its effect on the frequencies of
GWs. However, it is known for light waves that a relative motion not only changes the
frequencies but also the brightness of the source, the latter of which is
called the ``beaming effect''. Here we investigate such an effect for
GWs and find that the observed amplitude of a GW signal, unlike the
behavior of light, is not a monotonic function of the relative velocity and
responds differently to the two GW polarizations. We attribute the difference
to a rotation of the wave-vector, as well as a reorientation of
the GW polarizations. We find that even for
velocities as small as $0.25\%$ of the speed of light, ignoring the aforementioned
beaming effect could induce a systematic error
that is larger than the designated calibration accuracy of LIGO. 
This error could lead to an incorrect estimation of the distance and orbital
inclination of a GW source, or result in a spurious signal that appears
to be incompatible with general relativity.
\\[2mm]\indent
\end{abstract}

\maketitle

\section{Introduction}

The direct detection of gravitational waves (GWs) by the ground-based
observatories LIGO and Virgo~\cite{ligo_2018a,ligo_2018b} has led in the last few
years to the advent of data-driven gravitational wave astronomy. A key
ingredient in the detections is the success of numerical relativity to
reproduce the inspiral, merger, and ringdown waveforms of compact binaries,
such as binary black holes (BBHs,
e.g.~\cite{pretorius_2005,campanelli_lousto_2006,baker_centrella_2006}).  In
alliance with various inspiral modelling methods
(e.g.~\cite{buonanno_damour_1999,buonanno_pan_2007,ajith_hannam_2011,santamaria_ohme_2010}),
these waveform templates have allowed us to extract from the observed signals
physical parameters for the binaries, such as the masses of the compact objects
and their spins.

Although the templates are computed in a coordinate system where the
center-of-mass (CoM) of the source is at rest, observations are conducted in a
different frame, where the detectors are at rest.  The two frames, in
general, are not equal because astrophysical objects move relative to
us. As a result, the computed and the observed waveforms may differ.
Indeed, recent works considered the astrophysical scenarios in which the
sources are moving at a non-relativistic velocity
and showed that if the velocity varies with time, the acceleration could
induce a detectable difference in the observed waveform, by shifting the GW
frequency (the Doppler
effect~\cite{bonvin_caprini_2016,gerosa_moore_2016,meiron_kocsis_2016,inayoshi_haiman_2016,chamberlain_moore_2019})
or changing the way the high-order GW modes
interact~\cite{calderon-bustillo_clark_2018}.

The difference should be the most prominent when the relative velocity
approaches the speed of light, $c$. This case, however, is not considered
in the previous studies.  The main reason is that gaining a relativistic
velocity is considered difficult for BBHs in normal astrophysical
environments~\cite{ligo_2016a,amaro-seoane_chen_2016}.

This conventional view, however, is no longer complete. Recent studies
showed that BBHs could merge more rapidly in the center of a galaxy, especially in
the presence of a supermassive black hole (SMBH). The mergers are caused by
either the tidal perturbation by the SMBH
\cite{antonini_perets_2012,prodan_antonini_2015,stephan_naoz_2016,van-landingham_miller_2016,liu_wang_2017,petrovich_antonini_2017,bradnick_mandel_2017,hoang_naoz_2017,arca-sedda_gualandris_2018,fragione_grishin_2018}
or the hydrodynamical friction against an accretion disk, if the SMBH is in an
active galactic nucleus (AGN, e.g.
\cite{syer_clarke_1991,bellovary_mac-low_2016,bartos_kocsis_2017,stone_metzger_2017,mckernan_ford_2018}).  In particular, as
we found recently, a small fraction of the mergers could happen within a distance of ten
Schwarzschild radii of a SMBH \cite{chen_li_2017,chen_han_2018,han_chen_2018}.
Consequently, the CoM of the BBHs would move at a velocity of ${\cal
O}(c/\sqrt{20})$ relative to a distant observer.

If BBHs gain such a relativistic velocity, the power of the GWs, when viewed in
the rest frame of the detector, should, in principle, appear beamed in the
direction of the motion.  Such a ``beaming effect'' is well known for light
waves~\cite{johnson_teller_1982,jackson_2009} but not as well understood for GWs.
Theoretically, the standard equation for the generation of GWs assumes a slow
motion for the source \cite{peters_mathews_1963,peters_1964}, and hence is not easily generalizable to account for a
relativistic velocity.  A self-consistent treatment of
the problem should, instead, consider a source which extends into its own wave
zone, but the resulting equation is difficult to solve, except for a few
special cases \cite{press_1977}.

One possibility of simplifying the calculation is to return to the GW
strain previously computed in the rest frame of the source and Lorentz
transform it into the rest frame of the detector.  This operation, although
theoretically appropriate, may result in an observational issue, especially for
BBHs, because the Lorentz transformation changes the
apparent direction of the GW polarization
\cite{thorne_1987}. This direction  is used to infer the orientation of the orbit
of a BBH \cite{sathyaprakash_schutz_2009}, which is particularly useful to
constrain the orbital precession, black-hole spin, or alternative gravity
theories \cite{gair_vallisneri_2013}.

To overcome the above difficulties, we take a different approach towards
addressing the beaming effect of GWs.  We stay in the rest frame of the source
and investigate the response of a moving detector to the GW background. The
advantage of this approach is that one can use our response function
(equivalent to an ``antenna pattern'') to extract the GW waveform in the source
frame. This waveform can be compared directly with the templates from
numerical relativity to infer correct physical parameters of the source,
including the orientation of the orbit. Throughout this paper, we use $G=c=1$.

\section{The basic scenario}\label{sec:basic}

We first revisit the textbook example in which a pulse
of light is sent out by an emitter, reflected back by a mirror, and we measure
the duration of the round trip of the light (e.g.  \cite{schutz_2009}). This
example has the advantage of laying bare the fundamentals of
interferometric techniques as used in real detectors such as LIGO/Virgo \cite{ligo_2015} 
and the Laser Interferometer Space Antenna 
(LISA~\cite{danzmann_2017}). However, the textbook formulae are derived
assuming that there is no relative motion between the detector and the CoM of the source. This assumption no longer holds in our problem, and hence in this section we
generalize the formulae.

We choose a coordinate system $(t,\,x,\,y,\,z)$ at rest with respect to the CoM
of the GW source. This allows us to adopt the standard formulae for GW
radiation \cite{peters_mathews_1963,peters_1964}. Furthermore, we (i) expand
the spacetime metric to linear order, (ii) adopt the
\textit{transverse-traceless} (TT) gauge condition~\cite{misner_thorne_1973}, and (iii) set the wave-vector $\bs{k}$ of the GW in the $z$-direction and the $+$-polarization in the direction of the
$x$, $y$ coordinates. With these standard considerations, the spacetime metric
far away from the GW source reduces to
\begin{equation}\label{eq:gw}
g_{\mu\nu} = \eta_{\mu\nu} + h_{\mu\nu},
\end{equation}
where $\eta_{\mu\nu}:=\textnormal{diag}(1,\,-1,\,-1,\,-1)$ is the Minkowski metric and
$h_{\mu\nu}$ represents the GW. More specifically,
 the only non-vanishing components of $h_{\mu\nu}$ are
\begin{equation}\label{eq:pm}
h_{xx} = -h_{yy} = h_+^\omega(t,z) \fand h_{xy} = h_{yx} = h_\times^\omega(t,z).
\end{equation}

To simplify the problem, we focus on only one harmonic, where we denote its 
frequency as $\omega$ and its amplitude as $h_0$. Suppose the relative strength of the two polarizations are $\epsilon_+$ and $\epsilon_\times$, we then can write 
\begin{equation}\label{eq:hfun}
h_{+,\times}^\omega(t,z) = \epsilon_{+,\times}\,h_0\,e^{i(\omega t-kz)},
\end{equation}
where $i$ is the imaginary unit and $\omega=k$ in natural units.  We note that
in the standard case of a circular binary, $\epsilon_{+,\times}$ only depend on
the angle between the angular momentum of the orbit and the line of sight of
the observer~\cite{holz_hughes_2005}.  All the above quantities are
measured in the rest frame of the source.

Suppose the electromagnetic pulse sets off at the time $t_s$ from the
location of the emitter $\bs{p_E}(t_s)$, bounces back at the time $t_r$ by the
reflector located at $\bs{p_R}(t_r)$, and finally returns to the emitter at
$\bs{p_E}(t_e)$ at the time $t_e$. 
While in the textbook example the
above spatial vectors are constant,
in our case where both the emitter and the reflector are moving, these
vectors are functions of time. Given this difference,
we want to calculate the duration it takes,
i.e., $t_e-t_s$, for the pulse to finish the round trip. 

Since light travels along null geodesics, we use $g_{\mu\nu}\dd x^\mu \dd
x^\nu=0$ to compute the light travel time. Expanding the line element to linear
order of $h$, we derive
\begin{align}\label{eq:dtle}
\nn \dd t =& \sqrt{\dd x^2+\dd y^2+\dd z^2} - \frac12\frac1{\sqrt{\dd x^2+\dd y^2+\dd z^2}} \\
&\times\big[F_+(\dd x,\dd y)h_+^\omega(t,z) + F_\times(\dd x,\dd y)h_\times^\omega(t,z)\big],
\end{align}
where we have introduced the ``response patterns'' as
\begin{equation}\label{eq:ppc}
F_+(a,b) := (a^2-b^2) \fand F_\times(a,b) := 2ab.
\end{equation}
Integrating \eq{eq:dtle} using the boundaries
$\bs{p_E}(t_s)$, $\bs{p_R}(t_r)$, and  $\bs{p_E}(t_e)$, we find
\begin{align}\label{eq:ittf}
\nn t_e - t_s =& \int_{\bs{p_E}(t_s)}^{\bs{p_R}(t_r)}\Bigg(\sqrt{\dd x^2+\dd y^2+\dd z^2} \\
\nn &- \frac12\frac{P(\dd x,\dd y)}{\sqrt{\dd x^2+\dd y^2+\dd z^2}}h^\omega(t,z)\Bigg) \\
\nn &+ \int_{\bs{p_R}(t_r)}^{\bs{p_E}(t_e)}\Bigg(\sqrt{\dd x^2+\dd y^2+\dd z^2} \\
&- \frac12\frac{P(\dd x,\dd y)}{\sqrt{\dd x^2+\dd y^2+\dd z^2}}h^\omega(t,z)\Bigg).
\end{align}
In the last equation we have adopted for compactness
$h^\omega(t,z):=h_0\,e^{i(\omega t-kz)}$ and
\begin{equation}\label{eq:pp}
P(a,b) := \epsilon_+ F_+(a,b) + \epsilon_\times F_\times(a,b).
\end{equation}

Our calculation of the integrals in \eq{eq:ittf} involves a linear parameterization
of the geodesic of the light. For example, to  
calculate the first integral in \eq{eq:ittf}, which we denote as $I_1$,
we first notice that the arm length $L$ of LIGO or LISA is much shorter
than the corresponding GW wavelength $\lambda$,
i.e., we have $L/\lambda\ll1$ (see e.g. \cite{ligo_2015,danzmann_2017}). 
Consequently, we can linearize the geodesic as
\begin{equation}\label{eq:pcs}
\left(\begin{array}{c} t \\ \bs{\phi} \end{array}\right) = \left(\begin{array}{c} \beta^0 \\ \bs{\beta} \end{array}\right)\xi + \left(\begin{array}{c} \gamma^0\\ \bs{\gamma} \end{array}\right),
\end{equation}
where $\bs{\phi}$ are the three spacial components of the null geodesic of the light. Assuming that the two boundaries of the integration correspond to two values
$\xi_a$ and $\xi_b$, we can write
\begin{subequations}\label{eqs:pcsb}
\begin{align}\label{eq:pcsb1}
\left(\begin{array}{c} t_s \\ \bs{p_E}(t_s) \end{array}\right) =& \left(\begin{array}{c} \beta^0 \\ \bs{\beta} \end{array}\right)\xi_a + \left(\begin{array}{c} \gamma^0\\ \bs{\gamma} \end{array}\right), \\ \label{eq:pcsb2}
\left(\begin{array}{c} t_r \\ \bs{p_R}(t_r) \end{array}\right) =& \left(\begin{array}{c} \beta^0 \\ \bs{\beta} \end{array}\right)\xi_b + \left(\begin{array}{c} \gamma^0\\ \bs{\gamma} \end{array}\right).
\end{align}
\end{subequations}
From these conditions the parameters $\beta^0$, $\bs{\beta}$, $\gamma^0$ and $\bs{\gamma}$ can be determined as
\begin{subequations}\label{eqs:pcsv}
\begin{align}\label{eq:pcsv1}
\left(\begin{array}{c} \beta^0 \\ \bs{\beta} \end{array}\right) =& \frac1{\xi_b - \xi_a}\left(\begin{array}{c} t_r - t_s \\ \bs{p_R}(t_r) - \bs{p_E}(t_s) \end{array}\right), \\ \label{eq:pcsv2}
\left(\begin{array}{c} \gamma^0\\ \bs{\gamma} \end{array}\right) =& \left(\begin{array}{c} t_r \\ \bs{p_R}(t_r) \end{array}\right) - \left(\begin{array}{c} \beta^0 \\ \bs{\beta} \end{array}\right)\xi_b.
\end{align}
\end{subequations}
Applying the parametrization in \eq{eq:pcs}, the integral $I_1$ takes the form
\begin{align}\label{eq:ttip}
\nn I_1 =& \int_{\xi_a}^{\xi_b} \Bigg(\sqrt{\bs{\beta}^2} - \frac12\frac{P\left(\beta^1,\beta^2\right)}{\sqrt{\bs{\beta}^2}} \\
&\times h^{\omega}(\beta^0\xi+\gamma^0,\beta^3\xi+\gamma^3)\Bigg)\dd\xi.
\end{align}
It can be solved analytically using the mathematically tools prepared in \aref{sec:appa}.
Similarly, we can also solve 
the second integral in \eq{eq:ittf}. Summing up the two integrals, we find
\begin{align}\label{eq:me}
\nn t_e - t_s =& \bigg[\sqrt{(\bs{p_R}(t_r)-\bs{p_E}(t_s))^2} \\
\nn &+\frac12\bigg(\frac{P(x_R(t_r)-x_E(t_s),y_R(t_r)-y_E(t_s))}{\sqrt{(\bs{p_R}(t_r)-\bs{p_E}(t_s))^2}} \\
\nn &\times i\frac{h^\omega(t_r,z_R(t_r))-h^\omega(t_s,z_E(t_s))}{\omega(t_r-t_s)-k(z_R(t_r)-z_E(t_s))}\bigg)\bigg] \\
\nn &+ \bigg[\sqrt{(\bs{p_E}(t_e)-\bs{p_R}(t_r))^2} \\
\nn &+\frac12\bigg(\frac{P(x_E(t_e)-x_R(t_r),y_E(t_e)-y_R(t_r))}{\sqrt{(\bs{p_E}(t_e)-\bs{p_R}(t_r))^2}} \\
&\times i\frac{h^\omega(t_e,z_E(t_e))-h^\omega(t_r,z_R(t_r))}{\omega(t_e-t_r)-k(z_E(t_e)-z_R(t_r))}\bigg)\bigg].
\end{align}

One can verify the last equation by considering the textbook example in which
the detector is at rest relative to the GW source. In this case we can write
\begin{equation}\label{eq:geor}
\bs{p_E}(t) = \bs{0} \fand \bs{p_R}(t) = L\bs{\hat{p}},
\end{equation}
where $L$ is the arm length of the detector and
$\bs{\hat{p}}=(\hat{x},\hat{y},\hat{z})$ is a unit vector pointing from the
emitter to the reflector, both quantities defined in the rest frame of the
source when there is no GW. The problem can be further simplified because
of the following factors.
First, the
travel time of the light for the outbound and inbound trip can be approximated
by
\begin{equation}\label{eq:stt}
t_r - t_s = t_1(1 + \mathfrak{h}_1) \fand t_e - t_r = t_2(1 + \mathfrak{h}_2),
\end{equation}
where $t_1$ and $t_2$ denote the light travel times without GWs,
and $\mathfrak{h}_1$ and $\mathfrak{h}_2$ are of the order of $h$. Second, because $L/\lambda$ is small, we can expand $h$ around $t_s$.
To linear order (of $L/\lambda$ and $h$) the result is
\begin{subequations}\label{eqs:exps}
\begin{align}\label{eq:exps1}
h^{\omega}(t_r,L\hat{z}) \approx&~ h^{\omega}(t_s) + i(\omega t_1-kL\hat{z})h^{\omega}(t_s), \\ \label{eq:exps2}
h^{\omega}(t_e) \approx&~ h^{\omega}(t_s) + i\omega(t_1 + t_2)h^{\omega}(t_s),
\end{align}
\end{subequations}
where $h^\omega(t)$ is shorthand for $h^\omega(t,0)$. Third, when there is no relative motion the coordinate time and the proper time
of the emitter are the same, $L$ equals the arm length in the rest frame of the
detector $L_0$, and $t_1=t_2=L_0$. Under these circumstances, \eq{eq:me} reduces to
\begin{align}\label{eq:sttr}
\tau_e-\tau_s =& 2L_0\bigg(1 - \frac12\Big[F_+\left(\hat{x},\hat{y}\right)h_+^\omega(\tau_s)\nonumber\\
&+ F_\times\left(\hat{x},\hat{y}\right)h_\times^\omega(\tau_s)\Big]\bigg),
\end{align}
where $\tau_s$ ($\tau_e$) is the proper time at the emitter when the pulse
leaves (returns). This equation is equivalent to that derived in text books
(e.g.~\cite{schutz_2009}). The above result is coordinate-independent, because
it is a proper time measured by the same clock.  

\section{The effect of relative motion}

Now we consider the effect induced by a relative velocity of
$\bs{v} = (v_x,v_y,v_z)$ between the detector and the CoM of the source.
Since we choose a coordinate system where the GW source is at rest,
$h_{\mu\nu}$ has the same components as in
\eq{eq:pm}. The problem reduces to solving the light travel time between the
two ends of a moving detector. In this case, the response of the detector
will differ fundamentally from 
\eq{eq:sttr} because the spatial coordinates of the
emitter and reflector are no longer constant.

More specifically, their motion not only has a linear component, $\bs{v}t$, but
also a non-linear one caused by the perturbation of GWs. We derive this
latter component in \aref{sec:appb}, which is based on the free-fall
geodesic equations for the emitter and reflector up to linear order of $h$. The resulting spatial coordinates of the geodesics are 
\begin{subequations}\label{eqs:geom}
\begin{align}\label{eq:geome}
\bs{p_E}(t) =& \bs{v}t + \frac{i\bs{\alpha}h^{\bar{\omega}}(t)}{\bar{\omega}(1-v_z)}, \\ \label{eq:geomr}
\bs{p_R}(t) =& L\bs{\hat{p}} + \bs{v}t + \frac{i\bs{\alpha}h^{\bar{\omega}}(t,L\hat{z})}{\bar{\omega}(1-v_z)},
\end{align}
\end{subequations}
where $\bar{\omega}:=\omega(1-v_z)$ and
\begin{equation}\label{eq:alp}
\bs{\alpha} := \left(\begin{array}{c} \frac12v_xP\left(v_x,v_y\right) - (1-v_z)[\epsilon_+ v_x+\epsilon_\times v_y] \\ \frac12v_yP\left(v_x,v_y\right) - (1-v_z)[\epsilon_\times v_x-\epsilon_+v_y] \\ -\frac12(1-v_z)P\left(v_x,v_y\right) \end{array}\right).
\end{equation}
We note that $1-v_z$ enters the equations because
for the detector $\omega -kv_z$ is the rate
at which the GW phase changes.

\seqs{eqs:geom} indicate that the emitter and reflector wiggle as they advance
along their geodesics. This wiggling can be understood from the fact that a
four-velocity has always constant magnitude. Since GWs deform the metric, the
four-velocity has to rearrange to preserve its magnitude, which in turn changes
the direction of the trajectory. Moreover, \eq{eq:alp} shows that when
$v_x=v_y=0$, the wiggling effect vanishes even if $v_z$ is non-zero. This is
because in TT-gauge the $t$- and $z$-components are not deformed by the GW.

\subsection{The light travel time}

Knowing $\bs{p_E}(t)$ and $\bs{p_R}(t)$, we can use them in \eq{eq:me} and
derive the duration of the round trip for light. The
calculation resembles that without velocity but with three differences. (i)
The length of the arm, $L$, and the light travel times, $t_1$ and $t_2$, differ
from those in the previous paragraph by a coordinate transformation. (ii)
The wiggling of the emitter and reflector changes the proper distance that light
traverses. Effectively, this means the first-order terms in
\eq{eq:stt} contribute to the calculation of the first and third terms in
\eq{eq:me}. (iii) We want to derive the proper time of the emitter, not the coordinate time,
because the latter is coordinate-dependent.

To proceed, we first calculate the light-travel times without GWs, which
are
\begin{equation}\label{eq:ttagwm}
t_{1,2} = \gamma^2L\left(\frac1{\gamma(\theta)} \pm v\cos(\theta)\right),
\end{equation}
where $\gamma:=(1-v^2)^{-1/2}$ is the Lorentz factor, $\gamma(\theta):=(1-v^2\sin^2(\theta))^{-1/2}$, 
and $\theta$ is the angle spanned
by the relative velocity and the arm of the detector, as seen in the rest frame
of the source. The length of the arm in the source frame is
\begin{equation}\label{eq:lt}
L = \frac{\gamma(\theta)}{\gamma}L_0.
\end{equation}

Next, we calculate the four terms in \eq{eq:me}. For
the first and third terms, we simplify them by performing three steps: (i) replace
the times using the approximations specified in \eq{eq:stt}, 
(ii) Taylor expand the roots to linear order of $h$, and (iii) expand the $h$ around $t_s$ up to linear order of $L/\lambda$, like in \seqs{eqs:exps}. 
Executing these steps, we find
\begin{subequations}\label{eqs:ttwv1}
\begin{align}\label{eq:ttwv11}
\nn &\sqrt{(\bs{p_R}(t_r) - \bs{p_E}(t_s))^2} = t_1(1+\mathfrak{h}_1) - \frac{L}{\gamma(\theta)}\mathfrak{h}_1 \\
\nn &+\frac{h^{\bar{\omega}}(t_s)}{2\bar{\omega}}\left(\bar{\omega}-\frac{\omega L\hat{z}}{t_1}\right)\bigg[P(v_x,v_y)t_1 \\
&+\left(2P(\hat{x},\hat{y},v_x,v_y) + \left(\hat{z}+\frac1{\gamma(\theta)}\right)\frac{P(v_x,v_y)}{1-v_z}\right)L\bigg], \\ \label{eq:ttwv12}
\nn &\sqrt{(\bs{p_E}(t_e) - \bs{p_R}(t_r))^2} = t_2(1+\mathfrak{h}_2) - \frac{L}{\gamma(\theta)}\mathfrak{h}_2 \\
\nn &+\frac{h^{\bar{\omega}}(t_s)}{2\bar{\omega}}\left(\bar{\omega}+\frac{\omega L\hat{z}}{t_2}\right)\bigg[P(v_x,v_y)t_2 \\
&-\left(2P(\hat{x},\hat{y},v_x,v_y) + \left(\hat{z}-\frac1{\gamma(\theta)}\right)\frac{P(v_x,v_y)}{1-v_z}\right)L\bigg],
\end{align}
\end{subequations}
where we have applied the properties of $\bs{\alpha}$ derived in \aref{sec:appa} to simplify the results.

Now we consider the second and forth terms in \eq{eq:me}. For the first
parts of these two terms we approximate the times as before and then use the properties of $P$ in \aref{sec:appa} to simplify the result. Because of the $h$ in the later terms we only need to keep the
zeroth-order terms in the expansion. Therefore, we derive
\begin{subequations}\label{eqs:ttwv2}
\begin{align}\label{eq:ttwv21}
\nn &\frac{P(x_r(t_r)-x_e(t_s),y_r(t_r)-y_e(t_s))}{\sqrt{(\bs{p_R}(t_r)-\bs{p_E}(t_s))^2}} = \\
&~\frac{L^2}{t_1}P\left(\hat{x},\hat{y}\right) + 2LP\left(\hat{x},\hat{y},v_x,v_y\right) + t_1P\left(v_x,v_y\right), \\ \label{eq:ttwv22}
\nn &\frac{P(x_e(t_e)-x_r(t_r),y_e(t_e)-y_r(t_r))}{\sqrt{(\bs{p_E}(t_e)-\bs{p_R}(t_r))^2}} = \\
&~\frac{L^2}{t_2}P\left(\hat{x},\hat{y}\right) - 2LP\left(\hat{x},\hat{y},v_x,v_y\right) + t_2P\left(v_x,v_y\right).
\end{align}
\end{subequations}

For the second parts of the second and forth terms, we expand the $h$
analogous to \seqs{eqs:exps} and find that
\begin{subequations}\label{eqs:ttwv3}
\begin{align}\label{eq:ttwv31}
i\frac{h^{\omega}(t_r,z_r(t_r)) - h^{\omega}(t_s,z_e(t_s))}{\omega(t_r-t_s) - k(z_r(t_r)-z_e(t_s))} =& ~h^{\bar\omega}(t_s), \\ \label{eq:ttwv32}
i\frac{h^{\omega}(t_e,z_e(t_e)) - h^{\omega}(t_r,z_r(t_r))}{\omega(t_e-t_r) - k(z_e(t_e)-z_r(t_r))} =& ~h^{\bar\omega}(t_s).
\end{align}
\end{subequations}

Because we need $\mathfrak{h}_1$ and $\mathfrak{h}_2$ to complete the calculation,
we revisit \eq{eq:stt} and notice
that $t_r-t_s$ equals to the sum of the first and second terms in \eq{eq:me},
and that $t_e-t_r$ equals to the sum of the third and forth terms.
From these two equations we find
\begin{subequations}\label{eqs:atd}
\begin{align}\label{eq:atd1}
\nn &\mathfrak{h}_1 = \bigg[\frac{1}{2(1-v_z)}P(v_x,v_y) - \frac{\gamma(\theta)\hat{z}L}{t_1(1-v_z)}P(\hat{x},\hat{y},v_x,v_y) \\
&-\hat{z}L\frac{\gamma(\theta)\hat{z}+1}{2t_1(1-v_z)^2}P(v_x,v_y) - \frac{\gamma(\theta)L}{2t_1}P(\hat{x},\hat{y})\bigg]h^{\bar{\omega}}(t_s), \\ \label{eq:atd2}
\nn &\mathfrak{h}_2 = \bigg[\frac{1}{2(1-v_z)}P(v_x,v_y) - \frac{\gamma(\theta)\hat{z}L}{t_2(1-v_z)}P(\hat{x},\hat{y},v_x,v_y) \\
&-\hat{z}L\frac{\gamma(\theta)\hat{z}-1}{2t_2(1-v_z)^2}P(v_x,v_y) - \frac{\gamma(\theta)L}{2t_2}P(\hat{x},\hat{y})\bigg]h^{\bar{\omega}}(t_s).
\end{align}
\end{subequations}

Using $t_e-t_s = t_1(1+\mathfrak{h}_1) + t_2(1+\mathfrak{h}_2)$, \eqs{eq:ttagwm}{eq:lt},
as well as the properties of $P$ in \aref{sec:appa}, we find
\begin{align}\label{eq:cttctr}
\nn t_e - t_s =& 2\gamma L_0\bigg(1  + \frac12\bigg[\frac{P(v_x,v_y)}{(1-v_z)} - \left(\frac{\gamma(\theta)}{\gamma}\right)^2 \\
&\times P\left(\hat{x}+\slashed{v}_x\hat{z},\hat{y}+\slashed{v}_y\hat{z}\right)\bigg]h^{\bar{\omega}}(t_s)\bigg),
\end{align}
where $\slashed{v}_{x,y}:=v_{x,y}/(1-v_z)$.

The last equation is coordinate-dependent because it is expressed in coordinate time. 
To get a coordinate-independent expression, we use the following relation between the coordinate time and the proper time of the emitter (see \aref{sec:appc}):
\begin{equation}\label{eq:tptct}
t_e - t_s = \gamma(\tau_e - \tau_s) + \gamma L_0\frac{P\left(v_x,v_y\right)}{1-v_z}h^{\gamma\bar{\omega}}(\tau_s).
\end{equation}
The term $\gamma\bar{\omega}$ in the last equation
is the Doppler-shifted frequency, which a moving detector will perceive. Moreover, 
we have used the relationship $h^{\bar{\omega}}(t_s)=h^{\gamma\bar{\omega}}(\tau_s)$ 
between the GW amplitudes in different frames,
which is accurate up to linear order of $h$.
Finally, we find that the light-travel time, which a clock fixed at the emitter
would measure, is
\begin{align}\label{eq:ttpt}
\nn \tau_e-\tau_s =& 2L_0\bigg(1 - \frac12\Big[\wt{F}_+\left(\bs{\hat{p}},\bs{v}\right)h_+^{\gamma\bar{\omega}}(\tau_s) \\
&+\wt{F}_\times\left(\bs{\hat{p}},\bs{v}\right)h_\times^{\gamma\bar{\omega}}(\tau_s)\Big]\bigg),
\end{align}
where $\wt{F}_+$ and $\wt{F}_\times$ are two new response patterns, i.e.,
\begin{equation}\label{eq:vpp}
\wt{F}_{+,\times}\left(\bs{\hat{p}},\bs{v}\right) := \left(\frac{\gamma(\theta)}{\gamma}\right)^2F_{+,\times}\left(\hat{x}+\slashed{v}_x\hat{z},\hat{y}+\slashed{v}_y\hat{z}\right).
\end{equation}

\subsection{Velocity dependent response patterns}

The response patterns in \eq{eq:vpp} differ in many ways from the classic ones
in \eq{eq:ppc}. This difference results from two independent effects. First,
the special-relativistic contraction of the arm in the rest frame of the source
results in a factor in front of $F_{+,\times}$.  The Lorentz factors enter as
quadratic terms because the response patterns are quadratic equations of the
length. Second, the general-relativistic wiggling of the emitter and reflector
relative to the GW source leads to the additional terms in the arguments of the
$F_{+,\times}$.

There are some special cases in which at least one of the above two effects vanishes.
(i) There is no relative motion. In this case, \eq{eq:vpp} reduces to the
so-called ``antenna patterns'' \cite{sathyaprakash_schutz_2009} and we recover
the classical light travel time in \eq{eq:sttr}. (ii) The arm is perpendicular
to the motion so that it is not contracted. (iii) There is a relative motion
but only in the $z$-direction, i.e.  $\slashed{v}_x=\slashed{v}_y=0$. As we
have discussed previously, the velocity four-vector in this case is not
affected by the GW so that the wiggling effect vanishes. (iv) The relative
motion is in an arbitrary direction, but the detector has a special orientation
such that the arm lies in the plane of the wave front, i.e. $\hat{z}=0$. In
this case, the emitter and the reflector encounter the same GW phase and hence
they also wiggle in phase. We note that this is the case considered in
Ref.~\cite{gerosa_moore_2016}, where the authors claimed that to linear order
there is no effect of relative motion on GW amplitude.

\begin{figure}[tbp] \centering \includegraphics[width=0.47\textwidth]{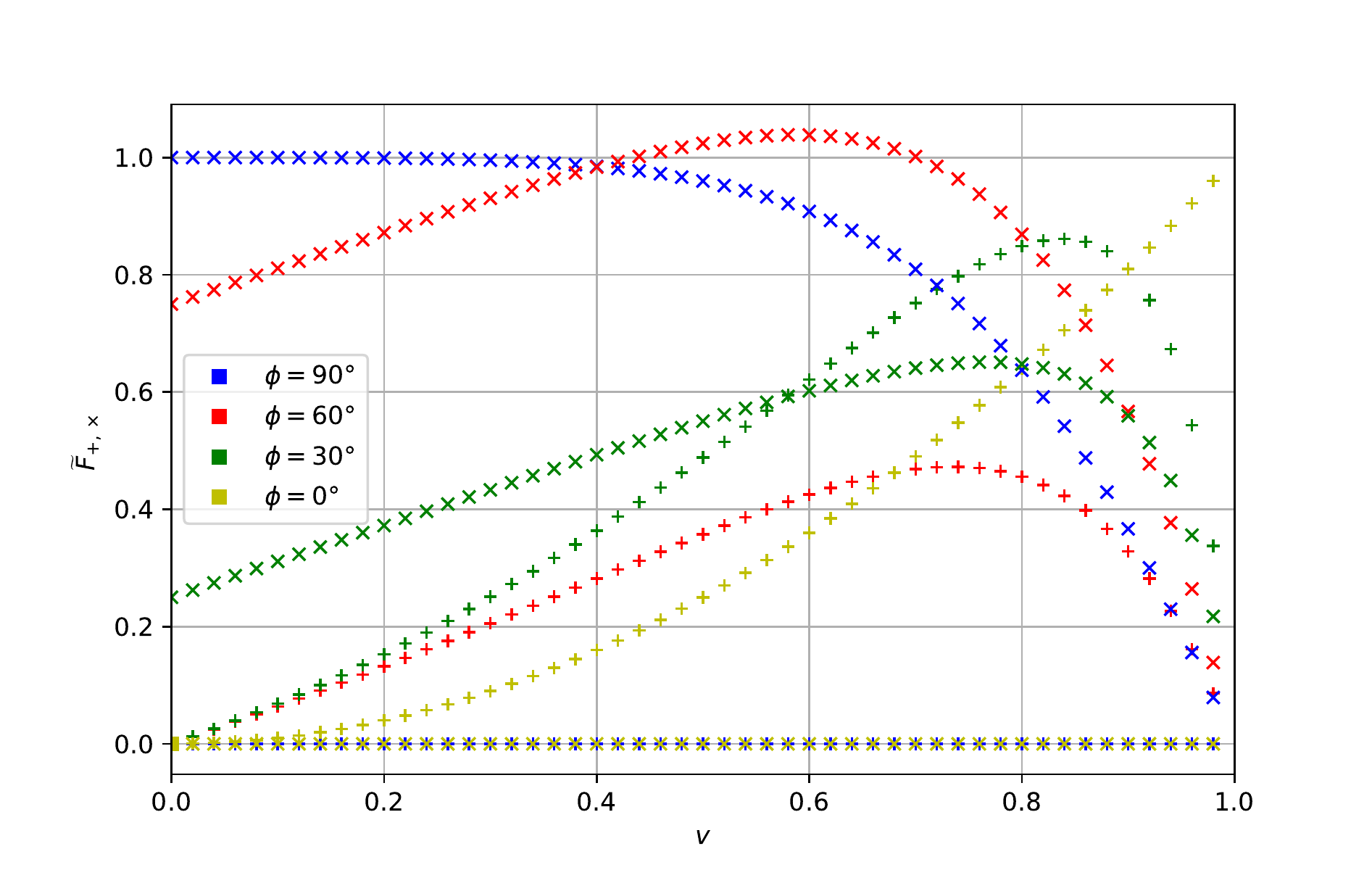}
\caption{
Magnitudes of the two response patterns as a function of the velocity $v$.
The plus symbols represent the pattern for the $+$-polarization and crosses for the $\times$-polarization.
In this example, the direction of the velocity is chosen to be along the $x$-axis. Both
the emitter and the reflector lie in the plane spanned by the $z$-axis and the angle bisector of
the $x$- and $y$-axes.
The angle $\phi$ denotes the orientation of the arm relative to the wave-vector, as seen
in the source frame.
} \label{fig:rep}
\end{figure}

To illustrate the behavior of our new response patterns, i.e., \eq{eq:vpp}, we
plot in \fref{fig:rep} a special case where the motion is in the $x$-direction,
i.e. $\bs{v}=(v,0,0)$. The arm formed by the emitter and reflector lies in the
plane spanned by the $z$-axis and the angle bisector of the $x$- and $y$-axes
and its position is described by the angle $\phi$ between the arm and the
wave-vector, as measured in the rest frame of the source. Without a velocity,
such an one-armed detector, by construction, would be blind to the
$+$-polarization. When there is a velocity, we find three remarkable features.
First, the patterns can either increase or diminish for an increasing velocity;
they are not monotonic functions of velocity.  Second, the dependence on
velocity is different for the two polarizations and for different
position angles of the arm. Third, the detector responds to the
$+$-polarization when it starts moving, except for the case $\phi=90^\circ$.
The third feature applies also to the $\times$-polarization if we start from a different
configuration in which the detector initially is blind to this polarization.

We emphasize that our approach allows us to find, in a unified way, two effects
on GWs due to relative motion. First, we recover the well-known Doppler effect
for GWs. Second, we find a different response of the detector to GWs, which will
fundamentally change the observable signals, as we will elaborate in the next
section.

\section{Detecting the beaming effect}

An interferometer (LIGO/Virgo/LISA) detects a passing GW by discerning,
effectively, the difference in the light travel times along the two arms of the same length, $L_0$,  pointing
in two different directions $\bs{\hat{p}_1}$ and $\bs{\hat{p}_2}$. This time
difference $\delta\tau$, which is a function of the GW phase, is the only observable in GW observations and closely related to the amplitude of a GW
signal, $h:= \delta\tau/L_0$~\cite{sathyaprakash_schutz_2009,schutz_2009}. Because it is measured by a single clock placed at the intersection of the two arms, it is a proper time and Lorentz invariant.

We now know that given a waveform template computed in the rest frame of the
source, one should use our response patterns, i.e. \eq{eq:vpp}, to derive
$h$ if the source is moving relative to the observer. In contrast,
if the observer is unaware of the relative motion and uses, instead, the
classic antenna patterns, i.e. \eq{eq:ppc}, to compute an amplitude
$h'$, it will be different from $h$.

To understand the difference between $h$ and $h'$, we 
study an ideal case in which the
source is a circular binary and its sky location, inclination, and distance
relative to the detector are known~\cite{peters_mathews_1963,peters_1964}. 
We can set up a coordinate system as described at the beginning of
the paper and compute the amplitudes of the signal using either \eq{eq:vpp} or
\eq{eq:ppc}. Moreover, we restrict our calculation to a representative example 
in which the relative velocity is in the $x$-direction. Furthermore, one arm is fixed in the $y$-direction,
i.e., $\hat{p}_1 = (0,1,0)$, while the other rotates in the $x$-$z$-plane, i.e., $\hat{p}_2 = (\cos(\theta),0,\sin(\theta))$. We choose this configuration because the fixed arm is not affected by the beaming effect while the rotating arm is. Under these circumstances, the two arms appear to be perpendicular to each other in both the source and the detector frame.

Finally, by varying the phase of GWs, we find that the maximum value for $h$ is
\begin{equation}\label{eq:dtttae}
h = h_0\epsilon_+\left[1 + \frac{1-v^2}{1-v^2\sin^2(\theta)}(\cos(\theta)+v\sin(\theta))^2\right],
\end{equation}
which an observer will take as the apparent magnitude of the GWs.
Alternatively, using the velocity-independent antenna patterns in \eq{eq:ppc}, we get
\begin{equation}
h' = h_0\epsilon_+\left[1 + \cos(\theta)^2\right].\label{eq:tauprim}
\end{equation}

\begin{figure}[tbp]
\centering
\includegraphics[width=0.47\textwidth]{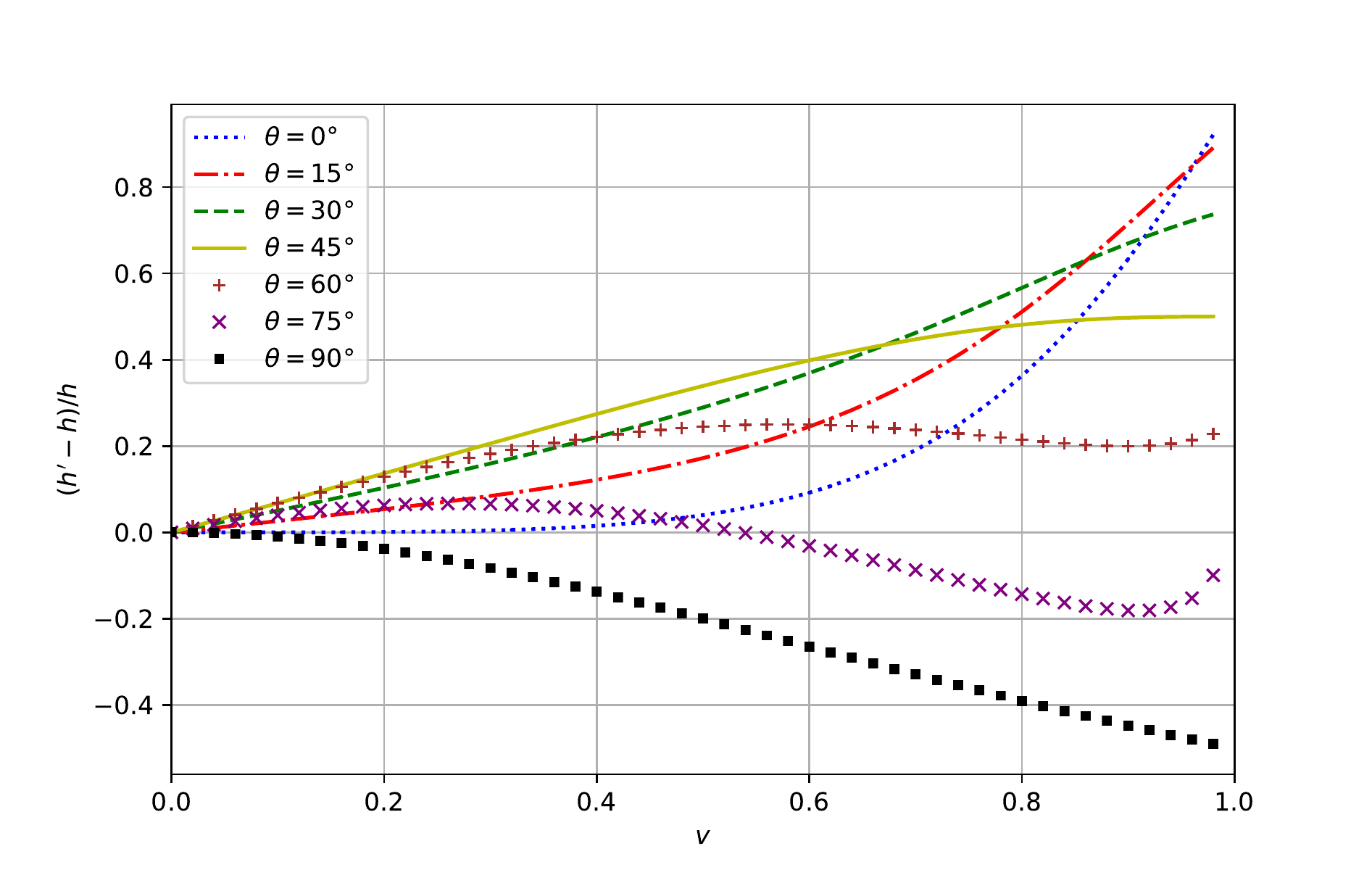}
\caption{Relative difference between
$h$ and $h'$ (the amplitudes of the GW signal) as a function of the velocity $v$.
Different curves refer to different orientations of the arm in the $x$-$z$ plane.
The amplitude of the signal $h$
is calculated using our response patterns which account for the relative motion,
while $h'$ is calculated using the classic antenna patterns which are 
velocity-independent. 
}
\label{fig:amp}
\end{figure}

\fref{fig:amp} shows the relative difference between $h$ and
$h'$. In general, the difference is a function of the velocity and the
orientation of the arms. Moreover, it is not always a monotonic function of the
velocity, and when the velocity is fixed the difference can be either positive
or negative depending on the orientation of the detector. We note that
in the case of light, where the observable is flux and not amplitude,
the beaming effect is, however, a monotonic function of velocity. Therefore, 
we find a fundamental difference between the beaming effect for light and that for
GWs.

\begin{figure}[tbp]
\centering
\includegraphics[width=0.47\textwidth]{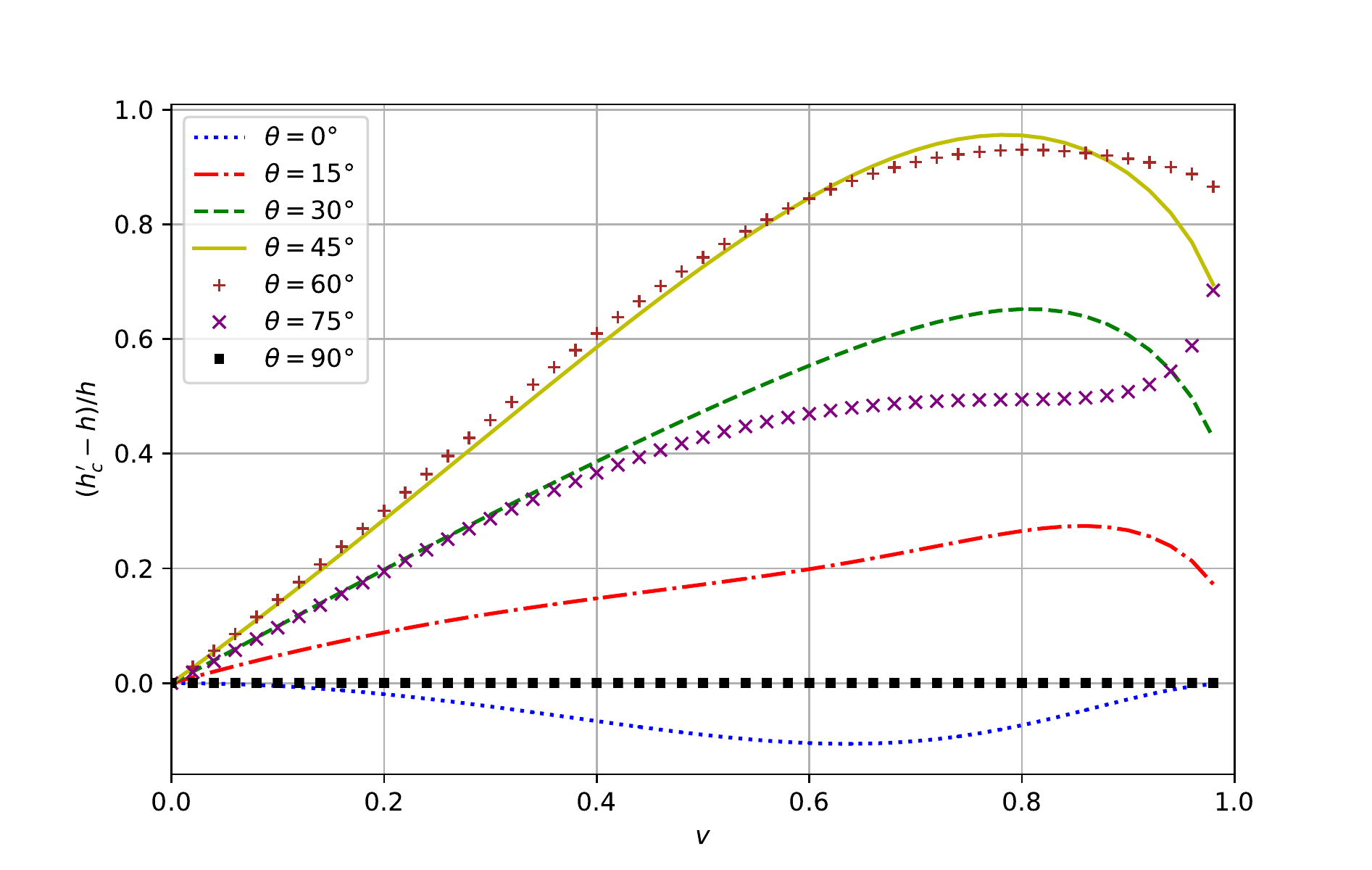}
\caption{The same as \fref{fig:amp} but in the calculation of $h'$ we correct the sky location of the source, as it is seen in the detector frame.
} \label{fig:ampc}
\end{figure}

To understand the cause of the difference shown in \fref{fig:amp}, we first
notice that for light the beaming effect can be fully explained by a Lorentz
transformation of the wave-vector~\cite{johnson_teller_1982}. For GWs, we can also
Lorentz transform the wave-vector into the detector frame, which,
effectively, changes the relative angle between the
wave-vector and our rotating arm. Placing this ``corrected'' angle in the classic antenna patterns 
while keeping the directions of the polarizations unchanged ($+$-polarization aligned with the
$x,y$-axes), we derive a new response of the detector:
\begin{equation}\label{eq:dtttaf}
h'_c = h_0\epsilon_+\left[1 + \left(\cos(\theta)/\gamma + v\sin(\theta)\right)^2\right].
\end{equation}
This result is different from the $h'$ derived in \eq{eq:tauprim}. The difference between 
$h$ and $h'_c$ is shown in
\fref{fig:ampc}. Although the difference vanishes in the case $\theta=90^\circ$,
it remains non-zero in the other cases. This result implies that the beaming
effect, as we have seen in \fref{fig:amp}, cannot be
attributed solely to the apparent change of the
sky location of the source. In fact, the apparent
directions of the polarizations also change~\cite{thorne_1987}. Given the
importance of the direction of GW polarization in the inference of the
orientation of the source, we will study in a future work the 
effect induced by a relative motion on GW polarization, as well as its astrophysical
implications.

\begin{figure}[tbp]
\centering
\includegraphics[width=0.47\textwidth]{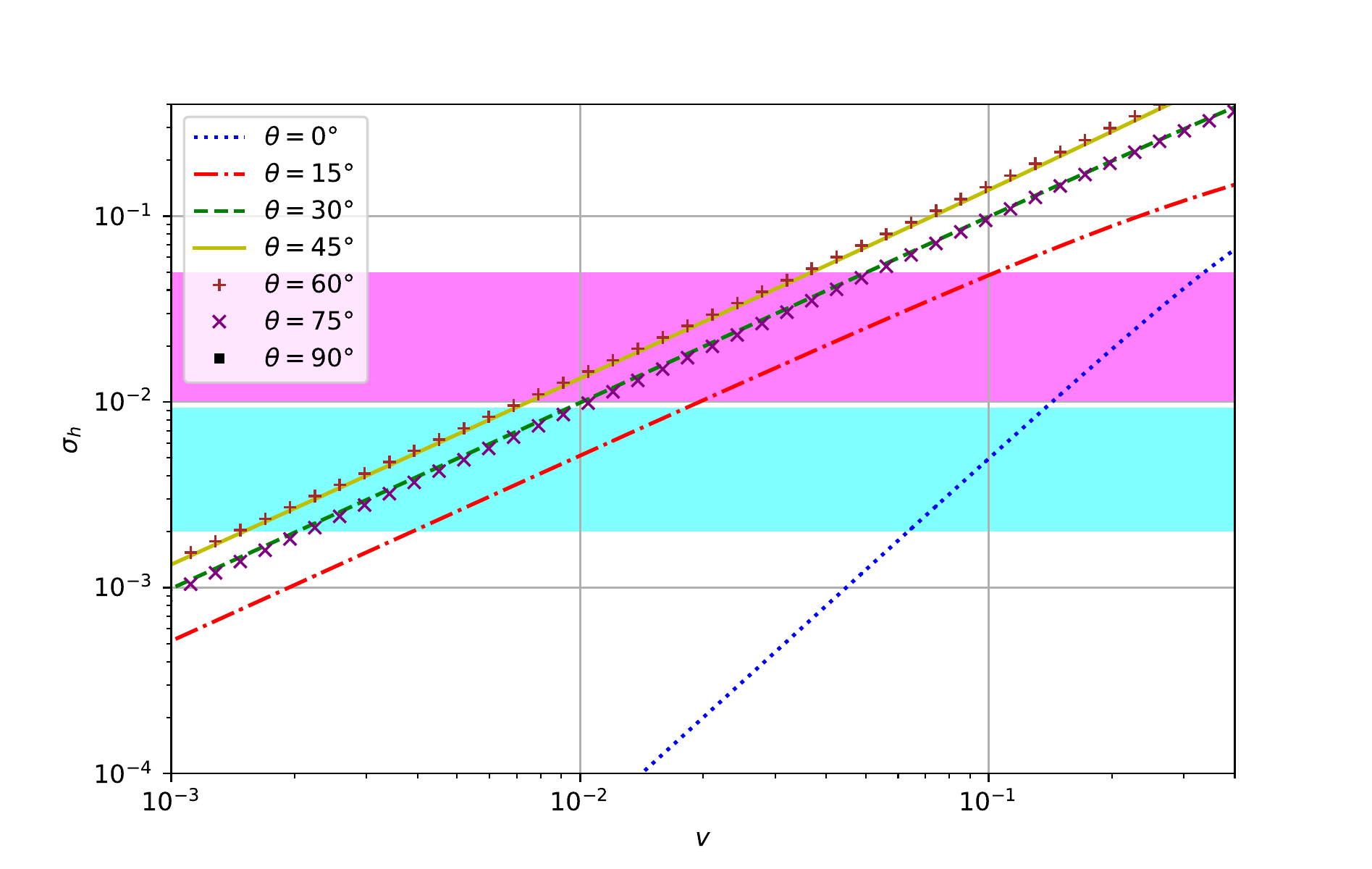}
\caption{Systematic error,
$\sigma_h:=|h'_c-h|/h$, in the prediction
of GW signals as a function of $v$. The lines are adopted from
\fref{fig:ampc}. The magenta shaded area shows the calibration accuracy in the
first and second LIGO observing runs. The cyan one shows the designated
calibration accuracy for the future runs.} \label{fig:errc} \end{figure}

So far, we have shown that ignoring the velocity of the source would result in
a systematic error in the prediction of the GW signal. \fref{fig:errc} shows
this systematic error for relatively low velocities. We find that it
is not negligible relative to the calibration accuracy of LIGO. For example,
when one of the arms is perpendicular and the other tilted by
$30^\circ-75^\circ$ relative to the direction of the velocity, a motion of
$(0.7-1.0)\%$ of $c$ would already induce a systematic error that exceeds the
best calibration accuracy of LIGO in the first and second observing
runs~\cite{cahillane_betzwieser_2017}.  For even higher velocities such as
$0.1\,c$~\cite{chen_li_2017,han_chen_2018,chen_han_2018}, the systematic error
could even exceed the upper limit of the  calibration accuracy of LIGO. In the future, LIGO
could further improve its calibration accuracy to
$(0.2-1.0)\%$~\cite{tuyenbayev_karki_2017,inoue_haino_2018}. According to the
same figure, even a velocities as small as $0.25\%$ of $c$ could cause a
significant error in the measurement of the amplitude of GWs.

\section{Conclusions}

Despite the seminal work of Isaacson~\cite{isaacson_1967} which shows that the
propagation of GWs is, in many aspects, similar to that of light, here we found that
the beaming effect for GWs, i.e., the responce of a detector to a moving source, is fundamentally
different.
First, the apparent amplitude of the GW signal, $h$, is not a monotonic function of the velocity. Second, the detector responds differently to the two
polarizations when the velocity changes. Third, and most
remarkably, the behavior of the signal can be explained partially, but not fully, by the
special-relativistic effects known for light, such as time dilation,
Doppler-shift, length contraction, and aberration.
The missing link is the re-orientation of the polarization
directions when there is a velocity. Because of the wide use of GW amplitudes to determine the distance
of the source \cite{schutz_1986} and the polarizations to infer the orbital
inclination \cite{sathyaprakash_schutz_2009} or test alternative gravity
theories \cite{gair_vallisneri_2013}, our results will have important
applications in the future for GW astrophysics and fundamental physics.

\begin{acknowledgments}

We thank Leor Barack, Cliff Will, and Carlos
Sopuerta for helpful discussions and suggestions. This work is supported by the
``985 Project'' of Peking University and the National Science Foundation of
China (No.~11873022, 11690023, and 11622546).  XC is partly supported
by the Strategic Priority Research Program of the Chinese Academy of Sciences,
Grant No. XDB23040100 and No. XDB23010200. PAS acknowledges support from the
Ram{\'o}n y Cajal Programme of the Ministry of Economy, Industry and
Competitiveness of Spain, as well as the COST Action GWverse CA16104.

\end{acknowledgments}

\appendix
\section{Some properties of $\bs{\alpha}$ and $P$}\label{sec:appa}

For $P$, defined in \eq{eq:pp}, we have:
\begin{subequations}\label{eqs:ppp}
\begin{align}\label{eq:ppp1}
P(\lambda a, \lambda b) =&~ \lambda^2P(a,b),
 \\ \label{eq:ppp2}
P(a+b,c+d) =&~ P(a,c) + 2P(a,c,b,d) + P(b,d),\end{align}
\end{subequations}
where
\begin{equation}\label{eq:ppgwd}
P(a,b,c,d) := \epsilon_+(ac - bd) + \epsilon_\times(ad + bc),
\end{equation}
which in turn fulfills
\begin{equation}\label{eq:ppp3}
P(\lambda a,\lambda b,c,d) = P(a,b,\lambda c,\lambda d) = \lambda P(a,b,c,d).
\end{equation}
For the $\bs{\alpha}$ defined in \eq{eq:alp}, $\bs{v}$ the relative velocity, and $\bs{\hat{p}}$ the direction of the arm, we find:
\begin{subequations}\label{eqs:alpr}
\begin{align}\label{eq:alprv}
\bs{v}\cdot\bs{\alpha} =& -\frac12(1-v_z)P\left(v_x,v_y\right) - \frac{P\left(v_x,v_y\right)}{2\gamma^2}, \\ \label{eq:alprl}
\bs{\hat{p}}\cdot\bs{\alpha} =& \frac{v\cos(\theta)-\hat{z}}2P\left(v_x,v_y\right) -(1-v_z)P\left(\hat{x},\hat{y},v_x,v_y\right).
\end{align}
\end{subequations}

\section{The geodesic of a moving particle}\label{sec:appb}

In this section we solve the geodesic of a test particle under the influence of
GWs.  We say the particle is initially at the point $\bs{p_0}=(x_0,y_0,z_0)$
and moves with a velocity $\bs{v}=(v_x,v_y,v_z)$, whereas the GW source is far
away and at rest. We apply the coordinate system and the metric introduced in Section~\ref{sec:basic}.

The geodesic equation for a point particle in a force free space
is~\cite{misner_thorne_1973}
\begin{equation}\label{eq:gect}
\ddot{x}^\alpha + \Gamma^\alpha_{\beta\gamma}\dot{x}^\beta\dot{x}^\gamma - \Gamma^0_{\beta\gamma}\dot{x}^\beta\dot{x}^\gamma\dot{x}^\alpha = 0,
\end{equation}
where we define $x^0 := t$ to write the geodesic equation in terms of the
coordinate time, the dot denotes the coordinate time derivative, and
$\Gamma^\alpha_{\beta\gamma}$ is the \textit{Christoffel symbol}. Moreover,
the geodesic of a massive particle has to fulfill the constraint
\begin{equation}\label{eq:gcct}
g_{\mu\nu}\dot{x}^\mu\dot{x}^\nu\left(\frac{\dd t}{\dd\tau}\right)^2 = 1,
\end{equation}
where $\tau$ is the proper time along the geodesic (see \aref{sec:appc}). For the
metric in \eq{eq:gw} the Christoffel symbols have the form
\begin{align}\label{eq:cs}
\nn \Gamma^\alpha_{\beta\gamma} =& -\frac12 i\omega h^{\omega}(t,z)\Big[\epsilon^\alpha_\beta (\delta^3_\gamma - \delta^0_\gamma) \\
&+ \epsilon^\alpha_\gamma (\delta^3_\beta - \delta^0_\beta) - \epsilon_{\beta\gamma} (\eta^{\alpha 3} - \eta^{\alpha 0})\Big],
\end{align}
where $\delta^\alpha_\beta$ is the \textit{Kronecker-Delta} and
$\epsilon_{\mu\nu}:=(h_{\mu\nu}|_{t,z=0})/h_0$.

We assume the velocity of the spatial coordinates $\bs{p}$ can be separated
into the initial velocity, $\bs{v}$, and a function $\bs{f}:=(f_x, f_y, f_z)$
of order $h$ describing the effect of the GW:
\begin{equation}\label{eq:sv}
\dot{\bs{p}} = \bs{v} + \bs{f}.
\end{equation}
Accordingly, $z(t) = z_0 + v_zt + g$, where $g$ is a function fulfilling
$\dot{g} = f_z$. Up to linear order in $h$ the geodesic equation is
\begin{equation}\label{eq:deg}
\ddot{\bs{p}} + i\omega\bs{\alpha}h^{\bar{\omega}}(t,z_0) = 0,
\end{equation}
where $\bs{\alpha}$ is introduced in \eq{eq:alp}. The general solution to
the differential equation is
\begin{equation}\label{eq:sge}
\bs{p}(t) = \bs{p_0} + \bs{v}t + \frac{i\bs{\alpha}h^{\bar{\omega}}(t,z_0)}{\bar{\omega}(1-v_z)},
\end{equation}
which also fulfills the constraint in \eq{eq:gcct}.

Considering the special case where the initial velocity vanishes, $\bs{\alpha}$
is also zero and the geodesic reduces to
\begin{equation}\label{eq:sgevv}
\bs{p}(t) = \bs{p_0},
\end{equation}
consistent with the classic notion that a particle at rest stays at rest in
the TT-gauge.

\section{Proper time along the geodesic}\label{sec:appc}

In this section we calculate the proper time along the geodesic derived in
\aref{sec:appb}. To linear order in $h$ it is related to the coordinate time as
\begin{equation}\label{eq:ptdf}
\dd\tau = \left(\frac1{\gamma} - \frac12\frac{P\left(v_x,v_y\right)}{\gamma(1-v_z)}h^{\bar{\omega}}(t,z_0)\right)\dd t.
\end{equation}

Integrating the left side from $\tau_a$ to $\tau_b$ and the right side from
$t_a$ to $t_b$ we find
\begin{align}\label{eq:pter}
\nn \tau_b - \tau_a =& \frac1\gamma(t_b-t_a) + \frac12\frac{P\left(v_x,v_y\right)}{\gamma\bar{\omega}(1-v_z)} \\
&\times i\big[h^{\bar{\omega}}(t_b,z_0) - h^{\bar{\omega}}(t_a,z_0)\big].
\end{align}
For zero velocity, we recover that $\tau_b-\tau_a$ equals $t_b-t_a$, i.e., the proper time is
the same as the coordinate time. For a non-vanishing velocity and neglecting
the terms linear in $h$, $\tau_b-\tau_a$ equals $(t_b-t_a)/\gamma$.
Rearranging \eq{eq:pter} we find
\begin{align}\label{eq:ctpt}
\nn t_b - t_a =& \gamma(\tau_b-\tau_a) - \frac12\frac{P\left(v_x,v_y\right)}{\bar{\omega}(1-v_z)} \\
&\times i\big[h^{\gamma\bar{\omega}}(\tau_b,z_0) - h^{\gamma\bar{\omega}}(\tau_a,z_0)\big],
\end{align}
where we have used $t=\gamma\tau$ to replace $t$ in the $h$.

To relate the coordinate time to the proper time along the
geodesic of the emitter we proceed as follows:
(i) we set $z_0=0$, (ii) $t_a$ and $t_b$ are replaced by $t_s$ and $t_e$,
respectively, and accordingly for the proper times, (iii) $h$ can be expanded
as in \seqs{eqs:exps}, and (iv) for the times in $h$ we can use
$t_s=\gamma\tau_s$. Finally, we find \eq{eq:tptct}.

\bibliographystyle{apsrev4-1.bst}
\bibliography{alebib}

\end{document}